\documentclass[aps,pra]{revtex4}
\newcommand{\eqref}[1]{(\ref{#1})}
\def\sech{{\rm sech\,}}
\begin{document}

\title{Bound State Wave Functions through the Quantum Hamilton-Jacobi
Formalism }
\author{S. Sree Ranjani$^{a} $\footnote{akksprs@uohyd.ernet.in},
K.G. Geojo$^{a} $\footnote{akksprs@uohyd.ernet.in}, A.K.
Kapoor$^{a} $\footnote{akksp@uohyd.ernet.in}, and P.K. Panigrahi
$^{b}$ 
\footnote{prasanta@prl.ernet.in}}
\address{$^a$ School of Physics, University of Hyderabad, Hyderabad 500 046, India\\
$^b$ Physical Research Laboratory, Navrangpura, Ahmedabad, 380
009, India}


\begin{abstract}
The bound state wave functions for a wide class of exactly
solvable potentials are found utilizing the quantum
Hamilton-Jacobi formalism of Leacock and Padgett. It is shown
that, exploiting the singularity structure of the quantum momentum
function, until now used only for obtaining the bound state
energies, one can straightforwardly find both the eigenvalues and
the corresponding eigenfunctions. After demonstrating the working
of this approach through a few solvable examples, we
consider Hamiltonians, which exhibit broken and unbroken phases of
supersymmetry. The natural emergence of the eigenspectra and the
wave functions, in both unbroken and the algebraically non-trivial
broken phase, demonstrates the utility of this formalism.
\end{abstract}

\maketitle

{\bf I  Introduction}\\

Amongst various approaches to quantum mechanical
problems,$^{\ref{sch}-\ref{suk}}$ the Hamilton-Jacobi
formalism,$^{\ref{lea},\ref{pad}}$ has been shown to be ideally suited
for determining the bound state energies of  exactly solvable
problems. For a lucid review of various formulations of quantum
mechanics and their relative merits and demerits, the reader is
referred to Styer {\it et. al}.$^{\ref{sty}}$ In classical mechanics,
the action-angle
variables are the most appropriate ones for the description of
the dynamical systems exhibiting periodic motion.$^{\ref{gol}}$ The
corresponding demonstration, for the quantum mechanical bound
state problems, is of recent origin. In 1984, Leacock and
Padgett$^{\ref{lea},\ref{pad}}$ defined an exact quantum action
variable through a quantum momentum function (QMF), whose quantization
condition
\begin{equation}
\frac{1}{2\pi} \oint p dx  = n \hbar,    \label{e1a}
\end{equation}
led to the bound state energies. The fact that only the residues
of the QMF are required for finding the eigenvalues, and the same
can be evaluated rather easily from the quantum Hamilton-Jacobi
(QHJ) equation$^{\ref{bh1},\ref{bh2}}$, explains the economy of the
QHJ formalism in finding the bound state spectra. In the limit
$\hbar \rightarrow 0 $ the QMF goes over to the classical momentum,
$\sqrt {2m(E-V)}$ and one can derive  the well-known WKB
quantization rule. The demonstration of the connection between the
supersymmetric WKB {\it approximation scheme}$^{ \ref{kharesref},\ref{k1}}$,
known to be exact for the so called shape-invariant potentials and
the {\it exact} Hamilton-Jacobi quantization condition further
clarified the working of various semi-classical quantization
methods$^{\ref{bh3}}$.

QHJ formalism is formulated in the complex plane, making it an
ideal tool for studying quasi-exactly
solvable$^{\ref{ushveridze},\ref{geo}}$ and recently discovered 
$PT$-symmetric potentials$^{\ref{bender}}$, since these are also
naturally studied in the same domain. It is well suited for
studying chaotic systems$^{\ref{gut}}$, where the periodic orbits
are known to play a significant role. In this connection it is
worth mentioning a recent demonstration of the close relationship
between the Kolmogorov-Sinai entropy and the action$^{\ref{par}}$,
both in classical and quantum domains. Before proceeding to the
analysis of the above mentioned non-trivial quantal problems, it
is necessary to develop a precise understanding of the procedure
for finding the wave functions from the quantum Hamilton-Jacobi
equation. Furthermore, in light of the close similarity between
the QHJ  and the WKB type quantization conditions and the well -
known difficulties one encounters in extracting the wave functions
in the semi - classical approaches, it is of deep interest to find
the same in the QHJ scheme.

The present work is devoted to the study of bound state wave
functions through the quantum Hamilton-Jacobi formalism. It is
shown that the singularity structure of the quantum momentum
function, which has been so far used only for finding the bound
state energies, can straightforwardly yield both the eigenvalues
and the corresponding eigenfunctions. In the earlier works of, two
of the present authors and those of Leacock and Padgett, the
singularity structure of the QMF and the boundary conditions it
needs to satisfy have been carefully studied. Since the same
potentials are being analyzed here, we use the singularity
structure of the QMF found earlier to determine the eigenstates.
We will also not deal with the details of the QHJ formalism and
refer the interested reader to earlier papers$^{\ref{lea},
\ref{pad}}$.

We first demonstrate the working of this approach through a number
of exactly solvable potentials, which have been studied
extensively in the literature through a number of approaches,
ranging from factorization method$^{\ref{inf}-\ref{gin}}$, path
integral techniques$^{\ref{kha},\ref{gro}}$, group theoretical
method$^{\ref{alh}}$ to SUSY operator method$^{\ref{khar}-\ref{suk}}$.
We then analyze Hamiltonians which exhibit broken and unbroken
phases of supersymmetry. The natural emergence of the eigenspectra
and the wave functions, in both the phases of supersymmetry,
demonstrates the utility of this formalism. It should be mentioned
that Hamiltonians exhibiting broken phase of supersymmetry are not
straightforwardly amenable to an algebraic treatment.

   The paper is organized as follows. We first give a brief
description of the QHJ formalism using the well-studied harmonic
oscillator problem as a prototype. We explicate the procedure to
obtain the bound state wave functions and its connection with the
singularity structure of the QMF.  This is an example of the confluent
hypergeometric class$^{\ref{grd}}$. We then proceed to the study of
Rosen-Morse and 
the Scarf - I potentials, belonging to the hypergeometric class. It is
interesting to note that, Scarf-I potential, although similar to
the Rosen-Morse, exhibits both broken and unbroken phases of
supersymmetry, for different ranges of potential
parameters$^{\ref{chu}-\ref{bh4}}$. The role of the proper
implementation of 
the boundary conditions, satisfied by the QMF in Hamilton-Jacobi
formalism, is clearly brought out in this section. It is shown
that, imposing the boundary conditions correctly leads
automatically  to the appropriate bound state wave functions, for
the broken and the unbroken phases of supersymmetry (SUSY).\\
\\
{\bf II  Harmonic Oscillator in the QHJ formalism}\\
\\
We begin this section with a brief summary of the QHJ formalism
for one dimensional potentials. As pointed out by Leacock and
Padgett, the QHJ equation,
\begin{equation}
p^{2}-i\hbar\frac{dp}{dx} = 2m(E-V(x)),     \label{e2}
\end{equation}
is a special case having the well-known Riccati equation,
\begin{equation}
A(x)p^{2} + B(x)p + C(x) + i \frac{dp}{dx} = 0.   \label{e2a}
\end{equation}
\eqref{e2} can be connected to the Schr$\ddot{o}$dinger equation
for a given potential $V(x)$ as
\begin{equation}
-\frac{\hbar^{2}}{2m}\frac{d^{2}\psi(x)}{dx^{2}}+V(x)\psi(x) =
E\psi(x) , \label{e1}
\end{equation}
provided the QMF $p$ is related to the wave function $\psi(x)$ by
\begin{equation} 
p = -i\hbar\frac{d}{dx}\ln\psi(x).    \label{e3}
\end{equation}
The above relation can be inverted to yield
$\psi(x)=\exp\left(\frac{i}{\hbar}\int pdx\right) $. We first
analytically continue $p$ to the complex plane by taking $x$ as a
complex variable. In what follows, the QMF plays a central role in
finding the bound state wave functions.  The knowledge of the
singularities of the QMF is sufficient to determine its form,
which consequently lead to the wave functions. The solutions of 
Riccati equation  has two types of singularities$^{\ref{inc}}$ , the 
fixed  and the moving singularities.  The fixed singular 
points  of $p(x)$ reflect the singularities of the potential and are energy
independent. Besides these, the QMF has other singular points called
the moving singularities of the type,
\begin{equation}
p \approx \frac{\hbar}{i}\frac{1}{x-a},     \label{e100}
\end{equation}
which cancel amongst the terms in the left hand side of
\eqref{e2}{}. It can be seen clearly from \eqref{e2}{}, that the
residue is given by $-i \hbar$. These poles, unlike the fixed
singularities, are dependent on energy and the initial conditions.
It is known that, for the solutions of the Riccati equation, a
moving singularity can only  be a pole$^{\ref{lea},\ref{pad},
\ref{inc}}$. It is a well- known fact that, the wave function of
the $n^{th}$ excited state will have $n$ zeros on the real line in
the classical region, which implies that the QMF will have $n$
poles. In all the solvable models studied within the QHJ approach,
no other moving singularity was found except the $n$ moving poles.
This observation has been very useful in solving for the bound
state energies$^{\ref{bh1},\ref{bh2}}$. In all the potentials
analyzed here, it has been assumed that, {\it the QMF has finite number of
singularities in the complex plane and that the point at infinity is an
isolated singular  point}.  In the present study we
find that, after a suitable change of variable followed by transformations on
$p$ to bring the QHJ equation to the form of \eqref{e2}{}, the
QMF is a rational function for the potentials studied here. The
knowledge of the location of poles and the corresponding residues
are sufficient to fix the form and to obtain the bound state wave
functions.

The QHJ equation can be used to obtain the residues at different
poles. Owing to the fact that the Riccati equation is quadratic,
one gets two answers for the residues. Hence, one needs a boundary
condition to select the correct value. The original boundary
condition
\begin{equation}
\lim_{\hbar \rightarrow 0}  p \rightarrow p_{cl},     \label{e3a}
\end{equation}
which was given by Leacock and Padgett$^{\ref{lea},\ref{pad}}$ can
be cast in several equivalent forms, which will be given as and
when they are used. We will use one of the convenient forms to
identify the correct residue.

    For the purpose of illustration, we start with the harmonic oscillator
$V(x) = 1/2(kx^2) $.
The Schr{\"o}dinger equation, with  $\xi=\alpha x$, $\lambda=
2E/\hbar \omega_c  ,\omega_c = \sqrt{k/m}$ and
$mk=\hbar^{2}\alpha^4 $,
\begin{equation}
\frac{\partial^{2}\tilde{\psi}(\xi)}{\partial\xi^{2}}+
(\lambda-\xi^{2})\tilde{\psi}(\xi) = 0 ,  \label{e4}
\end{equation}
has the corresponding QHJ equation in the form
\begin{equation}
p^2-i \frac{dp}{d\xi}=\lambda-\xi^2.   \label{e6}
\end{equation}
From the above equation it appears reasonable to assume that for
large $\xi$, $p$ goes as $\pm i\xi$. To fix the right behavior of
$p$ at infinity one can use the boundary condition \eqref{e3a} or
instead use the relation between $p$ and $\psi$
\begin{equation}
\psi(x) = \exp (i\int p dx).     \label{e7a}
\end{equation}
Demanding the square integrability of the wave function, the
correct behavior for large $\xi$ is seen to be
\begin{equation}
\lim_{\xi \rightarrow \infty} p \approx i\xi.    \label{e8}
\end{equation}
The QMF $p$ has $n$ moving poles on the real line with residue
$-i$. There are no fixed singularities in the finite domain in the
complex plane, since the potential is not singular. It is possible
to show that there are no other moving poles of the QMF in the
finite complex plane$^{\ref{lea},\ref{pad}}$, thus making $p$
analytic everywhere except at the $n$ moving poles. Therefore we
can write $p$ as
\begin{equation}
p(\xi)=\sum_{k=1}^{n}\frac{-i}{\xi-\xi_{k}} + i\xi+ Q(\xi),
\label{e9}
\end{equation}
where the summation term describes the sum of all the principal parts of the
individual Laurent expansion of $p$ around each moving pole. $Q(\xi)$ is an
entire function bounded for large $\xi$, with $i\xi$ describing the
behavior of $p$ at $\infty$. Thus from Liouville's theorem $Q(\xi)$ is
a constant $C$.
Hence the expression \eqref{e9} can be cast in a useful form
\begin{equation}
p(\xi) = -i\frac{P^{\prime}}{P} +i\xi + C \,     \label{e10}
\end{equation}
where $P$ is an $n^{th}$ degree polynomial $\prod_{k=1}^{n}(\xi -
\xi_{k})$ and is seen to satisfy 
\begin{equation}
P_{n}^{\prime\prime} - 2\xi P_{n}^{\prime} - 2CiP_{n}^{\prime}-(2i\xi C
+ C^2 - \lambda + 1)P_n = 0.    \label{e11}
\end{equation}
Equating the coeffecients of $\xi ^{n+1}$ and $\xi ^{n}$ to zero gives
\begin{equation}
C=0 ,\,\,  \lambda = 2n + 1  \label{e12}
\end{equation}
and hence $E_{n} = (n+\frac{1}{2})\hbar \omega_c$. With these values
\eqref{e11} becomes the Hermite equation and $P_n$ coincides with
Hermite polynomial $H_{n}(\xi)$ apart from an overall constant and
one obtains the known eigenfunctions
\begin{eqnarray}
\psi(\xi) = \exp\left(i \int p d\xi \right) \nonumber  \\ \label{e13}
          = \exp(-\frac{\xi^{2}}{2} +\ln(H_{n}(\xi))  \\ 
          = N H_{n}(\xi)\exp(-\xi ^2). \nonumber 
\end{eqnarray}
One can solve the radial part of hydrogen atom in the same way as the
above problem except that one encounters a fixed pole at the origin
along with the $n$ moving poles. Thus one needs to take this pole into account,
when one writes the behaviour of the QMF of the hydrogen atom in the
entire complex plane similar to \eqref{e12}. For more details one is
referred to $^{\ref{sree}}$ which is an earlier version of this paper.
The bound state wave functions for other solvable potentials, belonging to
confluent hypergeometric case, e.g., Morse and the radial
oscillator potentials can be obtained in the same way. In some
cases like Morse potential, one needs a suitable change of
variable, before solving the QHJ equation.\\
\\
{\bf III   Rosen-Morse Potential} \\
\\
This section is devoted to a study of solvable potentials
belonging to the hypergeometric class. Note that, in general one
does not come across potentials which are meromorphic in nature.
As mentioned earlier to make the QMF a rational function one needs to
do a suitable change of variable $ y = f(x)$. For a general
potential the QHJ equation 
\begin{equation}
p^2 - ip^{\prime} - (E - V(x)) = 0 \,   \label{e32}
\end{equation}
after a change of variable $ y = f(x)$ and introduction of $q$ by
$q(y)\equiv ip(x(y))$, becomes
\begin{equation}
q^{2}(y) + F(y)\frac{dq(y)}{dy} + E - \tilde{V}(y) = 0.
\label{e34}
\end{equation}
Here $F(y)$ equals the derivative $\frac{df(x)}{dx}$ expressed as
a function of $y$ and $\tilde{V}(y)=V(x(y))$. We see that
\eqref{e34}{} does not have the convenient form of \eqref{e2}{}.
To bring \eqref{e34}{} to the QHJ form we introduce $\chi$ by means of the
following equations.
\begin{equation}
q=F(y)\phi,\,\, \chi = \phi+ \frac{1}{2}\frac{d}{dy}(\ln F(y)),   \label{e36}
\end{equation}
and obtain
\begin{equation}
\chi^{2} + \frac{d\chi}{dy} + \frac{E-\tilde{V}(y)}{(F(y))^2} -
\frac{1}{2}\left(\frac{F^{\prime\prime}(y)}{F(y)}\right) +
\frac{1}{4}\left(\frac{F^{\prime}(y)}{F(y)}\right)^2
  =0      \label{e35}
\end{equation}

One can observe from \eqref{e35}{}, that the residues at the $n$
moving poles of $\chi$ will be unity. Therefore, we can make use of
\eqref{e35}{}, instead of \eqref{e32}{}, for any general potential
and proceed in the same way as for the harmonic oscillator.

For the Rosen-Morse potential, where the expression for the
potential is
\begin{equation}
V(x) = A^2 - A(A+\alpha )\sech ^{2}\alpha x    \label{e37}
\end{equation}
with $A>0$, we first change the variable to $y =\tanh(\alpha x)$. 
Using \eqref{e36}{}, one obtains the equation for $\chi$ as
follows :
\begin{equation}
\chi^2+\frac{d\chi}{dy}+\frac{E- A^{2} +\alpha^{2}}{\alpha^{2}
  (1-y^{2})^{2}}+\frac{A(A+\alpha)}{\alpha^{2}(1-y^{2})} = 0.    \label{e39}
\end{equation}
For the $n^{th}$ excited state, besides having $n$ moving poles with
residue equal to one, QMF will have fixed poles at $ y=\pm 1$. We
assume that there are no other singular points in the finite complex - y
plane. Thus proceeding in the same way as in the harmonic oscillator, one can
write $\chi(y)$ as the sum of the principal parts of the individual Laurent
expansions around different singular points, plus an analytic part which will
turn out to be a constant C. Thus
\begin{equation}
\chi(y)= \frac{b_{1}}{y-1}+\frac{b_{1}^{\prime}}{y+1}
+\frac{P^{\prime}}{P} +C ;  \label{e40}
\end{equation}
here  $b_{1}$ and $b_{1}^{\prime} $ are the residues at $y=\pm1$
respectively. One obtains the residues at these poles by doing a
Laurent expansion of $\chi(y)$ around each pole and substituting
it in the QHJ equation for $\chi(y)$. This leads to the values
\begin{equation}
b_1= \frac{1}{2}\pm\frac{\sqrt{A^{2}-E}}{2\alpha}\quad {\rm and}
\quad b_{1}^{\prime}=
\frac{1}{2}\pm\frac{\sqrt{A^{2}-E}}{2\alpha}. \label{e41}
\end{equation}
The ambiguity in the signs of the square roots can be eliminated
by applying the boundary condition \eqref{e1a}. Since we are
dealing with the SUSY potential, we can make use of the boundary
condition,
\begin{equation}
\lim_{ E\rightarrow 0}     q(y) = - W(y),    \label{e42}
\end{equation}
instead of the boundary condition \eqref{e3a} proposed by Leacock and Padgett,
with $W= A \tanh(\alpha x)$, the superpotential$^{\ref{coo}}$. Thus one 
obtains the correct residues as follows
\begin{equation}
b_1= \frac{1}{2} +\frac{\sqrt{A^{2}-E}}{2\alpha},\,\,b_{1}^{\prime}=
\frac{1}{2} +\frac{\sqrt{A^{2}-E}}{2\alpha}.     \label{e44}
\end{equation}
Substituting $\chi(y)$ with the right values
of $b_1$ and $ b_{1}^{\prime}$ in  \eqref{e39}{}, and for large $y$ comparing
the coefficients of $1/y$ and the coefficients of $1/y^{2}$, one gets
\begin{equation}
C = 0  ,\,\, E= A^2 -  (A-n\alpha)^2    \label{e45}
\end{equation}
respectively and is left with the differential equation
\begin{equation}
\frac{P^{\prime\prime}}{P}+\frac{2P^{\prime}y}{P}\left(\frac{\alpha+\sqrt{A^{2}-E}}{\alpha(y^{2}-1)}\right)+\frac{(\alpha+\sqrt{A^{2}-E})^{2}}{2\alpha^{2}(y^{2}-1)}-\frac{E-A^{2}+\alpha^{2}}{2\alpha^{2}(y^{2}-1)}-\frac{A(A+\alpha)}{\alpha^{2}(y^{2}-1)}  = 0.  \label{e47}
\end{equation}
Substitution of the energy eigenvalue in the above equation gives the Jacobi
differential equation for $P$
\begin{equation}
(1-y^2)P^{\prime\prime} - y(\mu +\nu +2)P^{\prime}+n(\mu +\nu +n+1)P = 0.   \label{e48}
\end{equation}
Thus $P$ is a  Jacobi polynomial represented by $P_{n}^{\mu,\nu}(y)$,
where $\mu = \nu = \frac{A}{\alpha}-n$. 
The bound state wave functions can be obtained by doing the change
of variable and using \eqref{e36}{}, in \eqref{e10}{} gives
\begin{equation}
\psi(y) = \exp\left(\int\chi(y)dy+\frac{y}{1-y^2}dy\right),
\label{e49}
\end{equation}
where $\chi(y)$ is
\begin{equation}
\chi(y) = \frac{\alpha+\sqrt{A^{2}-E}}{2\alpha(y-1)}
+\frac{\alpha+\sqrt{A^{2}-E}}{2\alpha(y+1)}
+\frac{P^{\prime}}{P}\, .    \label{e50}
\end{equation}
Thus the bound state wave function from \eqref{e49}{} turns out to
be
\begin{equation}
\psi_{n}(y)= N(1-y)^{\frac{\mu}{2}}(1+y)^{\frac{\nu}{2}}P_{n}^{\mu ,\nu }(y)  \label{e51}
\end{equation}
which matches with the known results $^{\ref{coo}}$ for $A$, real and
positive.\\
\\
{\bf IV    Scarf - I (trigonometric)}\\

There exist potentials$^{\ref{chu}-\ref{rd}}$ for which SUSY can be
exact or broken, depending on the range of potential parameters.
As mentioned earlier in the introduction, Bhalla {\it   et
  al.}$^{\ref{bh4}}$ have studied the eigenspectra of Scarf - I
potential. In this section, we study this potential and obtain the
expressions for the wave functions. The expression for the
potential is 
\begin{equation}
V_{-}(x) = - A^2 + (A^{2}+B^{2}-A\alpha \hbar)\sec^{2}\alpha x -
B(2A-\alpha \hbar)\tan\alpha x\sec\alpha x.   \label{e53}
\end{equation}
For this potential, SUSY is known to be  exact in the parameter range
\begin{equation}
(A-B)>0,\,\, (A+B)>0,    \label{e54}
\end{equation}
and broken in the range
\begin{equation}
(A-B)>0 ,\,\, (A+B)<0 .   \label{e55}
\end{equation}
Since broken SUSY is one of the possibilities, unlike the previous cases, we
cannot assume SUSY
to be exact and hence cannot use the superpotential to find the correct
residue at the fixed poles. We therefore, fall back upon the boundary
condition originally proposed by Leacock and Padgett$^{\ref{lea},\ref{pad}}$ 
\begin{equation}
\lim_{\hbar \rightarrow 0} p \rightarrow p_{cl}.     \label{e56}
\end{equation}
It turns out that imposing the boundary condition consistently for
different ranges of the parameters $A$ and $B$, automatically
gives rise to the correct wave functions and energy eigenvalues for
both broken and unbroken phases of SUSY. Retaining $\hbar$ with $(2m=1)$,
replacing $p$ with $-iq$ in the QHJ equation, one gets,
\begin{equation}
q^2 + \hbar \frac{dq}{dx}+E+ A^2 - (A^{2}+B^{2}-A\alpha \hbar)\sec^{2}\alpha x +
B(2A-\alpha \hbar)\tan\alpha x\sec\alpha x = 0.  \label{e57}
\end{equation}
Making a change of variable
\begin{equation}
y = \sin\alpha x,   \label{e58}
\end{equation}
and switching to $\chi$ in place of $q$ as explained previously, we get
\begin{equation}
\chi^{2}+\hbar \frac{d \chi}{dy}+\frac{y^{2}\hbar^{2}}{4(1-y^2)^{2}}
+\frac{E+A^{2}}{\alpha^{2}(1-y^{2})}
+\frac{\alpha^{2}\hbar^{2}-2(A^{2}+B^{2}-A\alpha\hbar)}{2\alpha^{2}(1-y^2)^2}
+\frac{B(2A-\alpha \hbar)y}{\alpha^{2}(1-y^2)^2} = 0.  \label{e59}
\end{equation}
Proceeding in the same way as before, we assume that there are no
singular points in the finite complex plane, other than the
fixed poles at $y=\pm1$ and the $n$ moving poles on the real line.
Therefore, we write $\chi(y)$ as
\begin{equation}
\chi(y)= \frac{b_{1}}{y-1}+\frac{b_{1}^{\prime}}{y+1}
+\hbar\frac{P^{\prime}}{P} +C,   \label{e60}
\end{equation}
where $b_{1}$ and $b_{1}^{\prime}$ are the residues at $y=\pm 1$ respectively.
The values of the residues $b_{1}$ and $b_{1}^{\prime}$ are obtained as in the
previous case.The values of $ b_{1}$ at $y =1$ are
\begin{equation}
b_{1} =\frac{(A-B)}{2\alpha}+\frac{\hbar}{4} \quad {\rm and} \quad
-\frac{(A-B)}{2\alpha}+\frac{3\hbar}{4} \, ;   \label{e61}
\end{equation}
The two values of the residue at $y=-1$ are
\begin{equation}
b_{1}^{\prime}=\frac{(A+B)}{2\alpha}+\frac{\hbar}{4} \quad {\rm
and} \quad -\frac{(A+B)}{2\alpha}+\frac{3\hbar}{4}. \label{e62}
\end{equation}
Different pairs of $b_{1}$ and $b_{1}^{\prime}$ give physically
acceptable solutions in the two phases of SUSY. We proceed to
obtain these solutions by taking the correct choice of the
residues using \eqref{e56}{}, the details of which are similar to
those in Bhalla ${\it et.al.}$$^{\ref{bh4}}$ \footnote{Due to 
typographical errors in the table 3 of$^{\ref{bh4}}$, the results do
not match the results obtained in this study.} The boundary
condition for this range of values of $A$ and $B$ implies, 
$ b_1>0 $  and $ b_{1}^{ \prime }>0 $ in the limit $ \hbar \rightarrow 0
$. \\

{\it Case:1 $ A+B>0,A-B>0$ }\\

For this particular range of parameters the right choice  of the residues is
\begin{equation}
b_{1} =\frac{(A-B)}{2\alpha}+ \frac{\hbar}{4}
,\,\, b_{1}^{\prime}= \frac{(A+B)}{2\alpha}+\frac{\hbar}{4}.    \label{e63}
\end{equation}
Thus $\chi$ in \eqref{e60}{} becomes
\begin{equation}
\chi(y)=\left(\frac{A-B}{2\alpha}+ \frac{\hbar}{4}\right)\frac{1}{y-1} +
\left(\frac{A+B}{2\alpha}+ \frac{\hbar}{4}\right)\frac{1}{y+1}+\frac{P^{\prime}}{P}  \label{e64}
\end{equation}
substitution of which in \eqref{e59}{} and putting $\hbar=1$ gives the energy eigenvalue
\begin{equation}
E_{n}=(A+n\alpha)^{2} -A^2 \, .   \label{e65}
\end{equation}
The equation for $P$ assumes the form
\begin{equation}
(1-y^2)P^{\prime\prime} +(\mu -\nu-(\mu +\nu+1)y) P^{\prime} +n
(n+\mu +\nu)P = 0,     \label{e66}
\end{equation}
where $\mu=\frac{A-B}{\alpha}$ and $ \nu = \frac{A+B}{\alpha}$.
The above differential equation is of the form of the Jacobi
differential equation and hence $P$ coincides with
$P_{n}^{\mu-1/2,\nu-1/2} $. The bound state wave function when
SUSY is exact, is
\begin{equation}
\psi_{n}(y)=N(1-y)^{\frac{\mu}{2}}(1+y)^{\frac{\nu}{2}}P_{n}^{\mu-\frac{1}{2},\nu-\frac{1}{2}}(y).  \label{e67}
\end{equation}
These wave functions correspond to the solutions of the unbroken phase
of SUSY.$^{\ref{coo}}$ \\ 

{\it Case:2  $A-B>0, A+B<0$}\\

For this range of the parameters, the right choice of the residues is
\begin{equation}
b_{1} =\frac{(A-B)}{2\alpha}+\frac{\hbar}{4}
,\,\, b_{1}^{\prime}=-\frac{(A+B)}{2\alpha}+\frac{3\hbar}{4}.    \label{e69}
\end{equation}
Substitution of $\chi$ with these values of $b_1$ and $b^{\prime}_{1}$ in
\eqref{e59}{} and putting $ \hbar = 1$ gives the energy eigenvalue as
\begin{equation}
E=\left(B-(n+\frac{1}{2})\alpha\right)^{2} - A^2,    \label{e70}
\end{equation}
and the differential equation for $P$ turns out to be
\begin{equation}
(1-y^2)P^{\prime\prime}+((\nu -\mu+1) -y(\nu+\mu+2))P^{\prime}+n(n+1+\mu +\nu )P = 0
\label{e71}
\end{equation}
and the bound state wave function is found to be
\begin{equation}
\psi_{n}(y)=N(1-y)^{\frac{\mu}{2}}(1+y)^{\frac{\nu}{2}-\frac{1}{2}}
P_{n}^{\nu+\frac{1}{2},\mu-\frac{1}{2}}(y)    \label{e72}
\end{equation}
where $\mu=\frac{A-B}{\alpha}$ $ \nu=\frac{A+B}{\alpha}$.
These bound state wave functions correspond to the broken phase of
SUSY of the Scarf-I potential. The Scarf-I potential can be
related to the P{\"o}schl-Teller-I potential by redefinition of
the potential parameters as
\begin{equation}
\alpha x \rightarrow \alpha x+\frac{\pi}{2} ,\,\, A=\delta + \beta
\,{\rm and}\, B=\delta - \beta.   \label{e74}
\end{equation}
With this redefinition of the parameters and then putting $
\alpha=1$, we see that the bound state wave functions in
\eqref{e72}{} matches with the bound state wave functions of the
broken SUSY phase of the P{\"o}schl-Teller-I potential given
in$^{\ref{rd}}$. Thus we see that the QHJ formalism  in one dimension 
gives the correct expressions of the bound state wave functions,
when there are different phases of SUSY. It may be remarked here
that, in the range $A-B<0, A+B<0$, SUSY is exact but the roles of
$H_{-}$ and $H_{+}$ are interchanged. In the range $A-B<0, A+B>0$
SUSY is again broken.The methods outlined above, has been applied
to many other potentials like the Morse, Eckart,
P{\"o}schl-Teller(hyperbolic) and Scarf-II (hyperbolic) and the
results agree with those given
in.$^{\ref{coo}}$  \\

{\bf V    Conclusions}\\

In conclusion, the singularity structure of the quantum momentum
function, in the complex domain, can be effectively utilized for
obtaining both eigenfunctions and eigenvalues in an efficient
manner. Besides illustrating the procedure to find the wave
functions, through the analysis of the exactly solvable problems,
we have clarified the role of an appropriate implementation of the
boundary condition, necessary to solve the quantum Hamilton-Jacobi
equation. The same played an important role in yielding the
eigenvalues and eigenfunctions of potentials exhibiting broken and
unbroken phases of SUSY. As has been pointed out earlier, a close
connection between Kolmogorov-Sinai entropy and the action has
been obtained recently, making use of a Hamilton-Jacobi approach
through Bohm's formulation $^{\ref{boh}}$. The QHJ formalism of
Leacock and Padgett and the above are not identical. Hence, a
deeper exploration of the above connection, as well as the
extension of the Hamilton-Jacobi formalism to intrinsically
non-separable systems, like Hamiltonians whose classical dynamics
reveal chaotic behavior, need thorough investigation. Although,
we have dealt with solvable systems here, one needs to extend this
approach, to non-solvable cases, like various anharmonic
oscillators. We would like to mention that, during the course of
our  study, we have analyzed  scattering states of the Rosen-Morse
potential. It was found, by assuming that QMF has finite number of
moving singularities, one can obtain the scattering state wave
functions, provided the potential parameter $A$ is equal to an
integer $n$, a special case when the potential is reflectionless.
The case when $A \neq n$ needs further study, as also a more
careful analysis of the general scattering problem. The continuous energy
functions, of interest for scattering, are currently under study.\\

\noindent {\bf Acknowledgments}\\

\noindent KGG would like to  thank UGC for the financial support.
AKK acknowledges useful discussions with S. Chaturvedi and V.
Srinivasan.


{\bf References}


\begin{enumerate}

\item {\label{sch}}L. I. Schiff, {\it Quantum Mechanics} ( McGraw-Hill, Singapore, 1986).

\item {\label{inf}} L. Infeld  and T. E. Hull, {\it Rev. Mod. Phys.} {\bf 23}
  21 ( 1951).

\item {\label{foc}} A. S. Focas,  {\it J. Math. Phys.} {\bf 21}(6), 1318 (1980).

\item {\label{gin}} J. N. Ginocchio, {\it Ann. Phys.} {\bf 152}, 203 (1984).

\item {\label{kha}} K. C. Khandekar  and S. V. Lawande,  {\it Phys. Rep.}
{\bf137} (1986) and the references therein.

\item {\label{gro}} C. Grosche  and F. Steiner, {\it Handbook of Feynmann's
 Path  Integrals} ( Springer-Verlag, New York, 1998).

\item {\label{alh}} Y. Alhasad, F. G. G{\"u}rsey and F. Iachello,  {\it
  Ann. Phys.} {\bf 148}, 346 (1983).

\item {\label{khar}} R. Dutt , A. Khare  and U. Sukhatme, {\it
    Am. J. Phys.} {\bf 56}, 163 (1988).

\item {\label{coo}} F. Cooper, A. Khare  and U. Sukhatme,   {\it Supersymmetry
    in Quantum Mechanics} ( World Scientific, Singapore, 2001) and references 
therein.

\item {\label{suk}} F. Cooper, A. Khare and U. Sukhatme, {\it Phys. Rep.} {\bf
    251}, 267 (1995).

\item {\label{lea}} R. A. Leacock  and M. J. Padgett  {\it
    Phys. Rev. Lett}. {\bf 50}, 3 (1983).

\item {\label{pad}} R. A. Leacock  and M. J. Padgett  {\it Phys. Rev.} D
  {\bf 28}, 2491 (1983).

\item {\label{sty}} D. F. Styer {\it et al.} {\it Am. J. Phys. }
{\bf 70} (3), (2002).

\item {\label{gol}} H. Goldstein, {\it Classical Mechanics}, II edition
  (Addison-Wesley, USA, 1994).

\item {\label{bh1}} R.S. Bhalla, A.K. Kapoor  and P. K. Panigrahi,  {\it
Am. J. Phys.} {\bf 65}, 1187 (1997).

\item {\label{bh2}} R. S. Bhalla, A. K. Kapoor  and P. K. Panigrahi,  {\it
Mod. Phys. Lett.} A, {\bf 12}, 295 (1997).

\item{\label{kharesref}} A. Comtet, A. Bandrauk  and D. Campbell  
{\it Phys. Lett.} B {\bf 150} 159; 

\item{\label{k1}} A.Khare  1985 {\it Phys. Lett.} B
{\bf 161} 131 (1985).

\item {\label{bh3}}  R. S. Bhalla, A. K. Kapoor  and P. K. Panigrahi, {\it
    Phys. Rev. A}  {\bf 54}, 951 (1996).

\item {\label{ushveridze}} A. Ushveridze, {\it Quasi-Exactly Solvable Models
in Quantum Mechanics} ( Inst. of Physics Publishing, Bristol, 1994) and
references therein.

\item {\label{geo}} K. G. Geojo, S. Sree Ranjani and A. K. Kapoor,  {\it
    J. Phys A : Gen. Math} {\bf 36} 4591 (2003);  quant-ph 0207036.

\item{\label{bender}} C. M. Bender, D. C. Brody and H. F. Jones, 
{\it Phys. Rev. Lett.} {\bf 89}, 270401 (2002) and references therein.

\item {\label{gut}} M. C. Gutzwiller,  {\it Chaos in Classical and Quantum
    Mechanics} (Springer-Verlag, New York, 1990).

\item {\label{par}} M. H. Partovi, {\it Phys. Rev. Lett.} {\bf 89}, 144101-1
  (2002).

\item {\label{grd}} I. S. Gradshteyn and I. M. Rhyzhik, {\it Table of
    Integrals, Series and Products} ( Academic Press, New York, 1965).

\item {\label{chu}} C. X. Chuan, {\it J. Phys. A: Math. Gen.} {\bf 23}, L659
  (1992).

\item {\label{ral}} Y. Ralchenko  and V. Semonov, {\it J. Phys. A:
    Math. Gen.} {\bf24}, L1305 (1992).

\item {\label{rd}} R. Dutt, A. Khare and U. Sukhatme, {\it Phys. Lett.} A {\bf
    174}, 363 (1993).

\item {\label{bh4}} R. S. Bhalla, A. K. Kapoor and P. K. Panigrahi, {\it
    Int. J. Mod. Phys.} A, {\bf 12}, No. {\bf 10} 1875 (1997).

\item{\label{inc}} E. L. Ince,  {\it Ordinary Differential
equation} (Dover Publication Inc, New York 1956).

\item{\label{sree}} S. Sree Ranjani, K. G. Geojo, A. K. Kapoor,
  P. K. Panigrahi, quant-ph/0211168 (unpublished).

\item {\label{boh}} D. Bohm, {\it Phys. Rev.} {\bf 85,} 166 (1952).

\end{enumerate}


\end{document}